\documentclass[jkps,twocolumn,fleqn,showpacs,showkeys]{revtex4}   

\usepackage{graphicx}
\usepackage{amssymb}
\usepackage{amsmath}
\usepackage{bm}
\usepackage{kotex}
\usepackage{times}
\usepackage{xcolor}

\newcommand{\be}{\begin{equation}}
\newcommand{\ee}{\end{equation}}
\newcommand{\bea}{\begin{eqnarray}}
\newcommand{\eea}{\end{eqnarray}}

\newcommand{\etal}{{\it et al.}}

\begin{document}
\setcounter{page}{1}

\title[]{Measurement of Tidal Deformability in the Gravitational Wave Parameter Estimation for Nonspinning Binary Neutron Star Mergers}
\author{Yong-Beom \surname{Choi}}
\email{ ybchoi@pusan.ac.kr}
\author{Hee-Suk \surname{Cho}}
\email{chohs1439@pusan.ac.kr}
\author{Chang-Hwan \surname{Lee}}
\email{clee@pusan.ac.kr}
\affiliation{Department of Physics, Pusan National University, Busan 46241, Korea}
\author{Young-Min \surname{Kim}}
\email{ymkim715@unist.ac.kr}
\affiliation{Department of Physics, Ulsan National Institute of Science and Technology, Ulsan 44919, Korea}

\date[]{}

\begin{abstract}
One of the main targets for ground-based gravitational wave (GW) detectors such as Advanced LIGO
(Laser Interferometer Gravitational wave Observatory) and Virgo is coalescences of neutron star (NS) binaries.  
Even though a NS's macroscopic properties such as mass and radius
have been obtained from electro-magnetic wave observations,
its internal structure has been studied mainly by using theoretical approaches.
However, with the advent of Advanced LIGO and Virgo,
the tidal deformability of a NS, which depends on the internal structure of the NS,
has been recently  obtained from GW observations.
Therefore, reducing the measurement error of tidal deformability 
as small as possible in the GW parameter estimation is important. 
In this study, we introduce a post-Newtonian (PN) gravitational waveform model 
in which the tidal deformability contribution appears from 5 PN order,
and we use the Fisher matrix (FM) method to calculate parameter measurement errors. 
Because the FM is computed semi-analytically using the wave function, 
the measurement errors can be obtained much faster than 
those of practical parameter estimations based on Markov Chain Monte Carlo method. 
We investigate the measurement errors for mass and tidal deformability by applying the FM to the nonspinning TaylorF2 waveform model. 
We show that if the tidal deformability corrections are considered up to the 6 PN order, 
the measurement error for the dimensionless tidal deformability  can be reduced to about 
$75 \%$ compared to that obtained by considering only the 5 PN order correction.
\end{abstract}



\maketitle

%=======	Intro		================================	
\section{INTRODUCTION}
The presence of gravitational waves (GWs) was predicted by Einstein in 1916 \cite{Einstein1916} and 1918 \cite{Einstein1918}. 
The strain of GWs, however, is extremely small; hence, numerous efforts over the past decades to detect real GWs have failed. 
Finally, the first GW signal, named as GW150914, was captured by the two Advanced LIGO 
(Laser Interferometer Gravitational wave Observatory) detectors in September  2015. 
During the first and the second observing runs of the network of Advanced LIGO and Virgo, a total of 10 GW signals 
from binary black hole mergers were detected \cite{Abbott16, Abbott16_2, Abbott17, Abbott17_2, Abbott17_3,GWCatalog_1,GWCatalog_2}. 
Eventually, a multi-messenger astronomy era began with the observation of a GW signal 
from a binary neutron star (NS) merger in 2017 \cite{Abbott17_4, Abbott17_5}. 
Because KAGRA is planning to join the third observing run \cite{Somiya12, Akutsu17}, 
the expected detection rate of binary NS mergers is 3.2 $\sim$ 9.2 times per year with the four-detector network \cite{Dominik15}.

One of the main targets of Advanced LIGO and Virgo is compact binary coalescences (CBCs) 
in which a compact binary
means a black hole-black hole, black hole-NS, or NS-NS binary.
In data analysis for GWs from CBCs, a signal can be distinguished from noise 
by matching the detector data with theoretical waveforms. 
Once a GW signal is identified in the detection pipeline, 
a more detailed analysis can be carried out in  the parameter estimation pipeline based on the Markov Chain Monte Carlo method.
While the main purpose of the GW detection pipeline is to decide whether the GW signal enters the data stream or not,
the parameter estimation pipeline seeks  source properties  such as the mass and the spin of the system.
The result of the parameter estimation analysis is given by probability density functions 
for the parameters considered, and this process typically requires a very long time 
and a large amount of computing resources.
On the other hand, if the incident GW signal is strong enough 
and buried in Gaussian noise, the measurement errors can be easily calculated by using the Fisher matrix (FM) method,
which is a semi-analytic approach used to approximate the accuracy of the parameter estimation.

Recently, the tidal deformability of a NS was measured from the GWs generated by the merger of a neutron star
binary \cite{Abbott17_4,GW170817EOS}.
This parameter describes the response of a NS to the external tidal field generated by its companion star.
In the current GW parameter estimations, the observational constraints on the NS radii are well estimated;
however, the measurement errors for tidal deformability are not small enough to precisely distinguish various equations of state (EOSs) \cite{Abbott17_4,GW170817EOS}.

In this work, we introduce a post-Newtonian (PN) waveform model, which is expressed in the frequency domain 
and is valid for the inspiral waveforms emitted from CBCs.
This model incorporates 3.5 PN order point-particle contributions and 5 PN and 6 PN tidal corrections.
We use the FM method and calculate the measurement errors of the mass and the tidal parameters for a nonspinning, equal-mass binary NS
assuming the Advanced LIGO detector sensitivity. 
We investigate how much the 6 PN tidal correction can improve the accuracy in measuring the tidal deformability. 

%================= GW PARAMETERS AND POST-NEWTONIAN APPROXIMATION (section1)	====================

\section{Post-newtonian Gravitational waveforms} 
The physical parameters used to define waveforms from CBC systems are divided into two groups. 
One group includes  intrinsic parameters such as mass and spin, 
which are directly related to the dynamics of a binary and the shape of the waveform. 
The other includes  extrinsic parameters such as luminosity distance, sky position, 
and inclination of the orbital axis. 
Unlike the intrinsic parameters, the extrinsic parameters only affect the wave amplitude; 
hence, they do not need to be considered in our analysis.
Generally, the correlation values between intrinsic and extrinsic parameters are much less than those between the intrinsic parameters \cite{Raymond12}.

The dynamics of CBCs in the relativistic region can be calculated using numerical simulations, 
which typically require a large amount of computing resources. 
However, in the region of slow orbital velocity ($v_0/c \ll1$) and weak gravitational field ($GM/(rc^2) \ll 1$), 
the post-Newtonian (PN) approximation is good enough
to describe the CBC dynamics and model the gravitational
waveforms from the CBC inspirals; hence, much fewer computing resources are required.

The frequency-domain PN waveform model obtained by using the stationary-phase approximation is called TaylorF2. 
The TaylorF2 waveform can be written as
\be
\tilde{h}(f) = A f^{-7/6} e^{i \psi(f)},
\ee
where $A$ is the wave amplitude consisting of the chirp mass ($M_c \equiv (m_1 m_2)^{3/5}/M^{1/5}$, where $M$ is the total mass) 
and the  extrinsic parameters. 
The phase $\psi(f)$ is expressed as the PN expansion 
\be \label{eq:phase_of_TaylorF2}
\psi(f) = 2\pi f t_c - \phi_c - \frac{\pi}{4} + \frac{3}{128\eta v^5} \bigg(\Psi_{\rm 3.5PN}^{\rm PP} + \Psi^{\rm Tidal} \bigg),
\ee
where $t_c$ and $\phi_c$ are the coalescence time and the coalescence phase, respectively, which can be chosen arbitrarily, $\eta \equiv m_1 m_2 / M^2$ is the symmetric mass ratio, and $v \equiv (\pi GM f)^{1/3}/c$ is the characteristic velocity. 
In the bracket, $\Psi_{\rm 3.5PN}^{\rm PP}$ is the point-particle contribution incorporated up to 3.5 PN order \cite{Buonanno09}, 
and $\Psi^{\rm Tidal}$ is the contribution by the quadrupolar tidal interaction, which is discussed in the next section.

 %================= TIDAL DEFORMABILITY (section2) =======================
\section{Tidal deformability} 
In a merging binary NS system, one NS can be deformed by the tidal field generated by the companion NS. 
To linear order, the quadrupole moment $Q_{ij}$ induced by the external quadrupolar tidal field 
$\mathcal{E}_{ij}$  is given by \cite{Flanagan08,Lackey12}
\be \label{eq:quadrupolar_deformation}
Q_{ij} = -  \lambda \mathcal{E}_{ij},
\ee
which can be decomposed with the symmetric traceless tensor $\mathcal{Y}^{lm}_{ij}$ 
defined by spherical harmonics $Y_{lm} (\theta,\phi)$ \cite{Hinderer09}. 
Only the leading contribution $l=2$ is included in our consideration; thus, the constant
$\lambda$ is the $l=2$ tidal deformability.
The quadrupole moment is related to density perturbation $\delta \rho$  as
\be
Q_{ij} = \int d^3x ~ \delta \rho(x) ~(x_i x_j - \delta_{ij}r^2 /3 ),
\ee
and $\mathcal{E}_{ij}$ can be expressed in terms of the external gravitational potential $\Phi_{\rm ext}$ as 
\be
\mathcal{E}_{ij} = \frac{\partial^2 \Phi_{\rm ext}}{\partial x^i \partial x^j}.
\ee
The tidal deformability is defined by \cite{Hinderer09,Flanagan08}
\be \label{eq:Love_number}
\lambda = \frac{2}{3G}k_2 R^5,
\ee
where $k_{2}$ is the Love number and $R$ is a radius of the NS. 
For a given NS mass, the tidal deformability can vary according to the EOSs \cite{Hinderer10}.

In a binary NS system, the tidal deformability contribution is given up to 6 PN order as \cite{Lackey15}
\be \label{eq:phase_by_tidal_deformability}
\Psi^{\rm Tidal}(f) = - \bigg[ \frac{39 \tilde{\Lambda}}{2}v^{10} + \bigg( \frac{3115\tilde{\Lambda}}{64} - \frac{6595 \sqrt{1-4\eta} ~ \delta \tilde{\Lambda}}{364} \bigg)v^{12} \bigg],
\ee
where the reduced tidal deformability $\tilde{\Lambda}$ and the asymmetric tidal correction $\delta \tilde{\Lambda}$ are defined as \cite{Lackey15}
\bea
\nonumber  \tilde{\Lambda} &=& 32 \frac{c^{10} \tilde{\lambda}}{G^4 M^5} \\ 
\nonumber  & =& \frac{16c^{10}}{13 G^4}\bigg(\frac{m_1 + 12m_2}{m_1}\lambda_1 + \frac{m_2 + 12m_1}{m_2}\lambda_2 \bigg) \\ 
\nonumber & =& \frac{8}{13}[ (1+7\eta - 31\eta^2)( \Lambda_1 + \Lambda_2 ) \\ 
 & +& \sqrt{1 - 4\eta}(1 +9\eta -11\eta^2)(\Lambda_1 - \Lambda_2)], \\
\nonumber \delta \tilde{\Lambda} &=& \frac{1}{2} \bigg[  \sqrt{1-4\eta} \bigg(1 - \frac{13272\eta}{1319} + \frac{8944\eta^2}{1319} \bigg)( \Lambda_1 + \Lambda_2 ) \\ \nonumber &+& \bigg( 1 - \frac{15910\eta}{1319} + \frac{32850\eta^2}{1319} + \frac{3380\eta^3}{1319} \bigg)( \Lambda_1 - \Lambda_2 ) \bigg], \\ 
&&
\eea
where $m_1 \ge m_2$, and $\Lambda_i = c^{10} \lambda_i / (G^4 {m_i}^5)$ is the dimensionless tidal deformability of a single star. 
The subscripts 1 and 2 indicate the individual NSs.
Because $\Lambda_1$ and $\Lambda_2$ are strongly correlated, $\tilde{\Lambda}$ and $\delta \tilde{\Lambda}$ are much more efficient in most analyses.
Typically, $\delta \tilde{\Lambda} / \tilde{\Lambda}$ is about 0.01;
hence, the contribution of $\delta \tilde{\Lambda}$ is negligible \cite{Favata14}. 
In this work, we assume $m_1 = m_2$ and $\Lambda_1 = \Lambda_2$; then, $\tilde{\Lambda} = \Lambda_1 = \Lambda_2$ 
and the $\delta \tilde{\Lambda} $ term vanishes in Eq.~(\ref{eq:phase_by_tidal_deformability}).

 %================= FM AND MEASUREMENT ERROR ESTIMATION (section3) =======================
\section{measurement error} 
Once a detection is made in the detection pipeline, 
the parameter estimation pipeline seeks the source parameters based on Markov Chain Monte Carlo method.
Although it is  guaranteed to converge, this method typically requires a long computational time.
On the other hand, 
the FM method has generally been used to predict accuracy of the parameter estimation
approximately for high signal-to-noise ratio (SNR) signals \cite{Cho14,Cho15-1} 
(for a general overview of the FM, refer to \cite{Vallisneri08}). 
The FM can be  calculated semi-analytically by using the post-Newtonian waveform;
therefore, the measurement errors can be obtained much faster than those of practical parameter estimation.

For a given theoretical waveform $h(t)$ and detector data stream $x(t)$ that can contain a real GW strain, the likelihood is determined by \cite{Cutler94, Finn92}
\be \label{eq:Likelyhood_relation}
 L(x|\theta) = \exp[-\langle x-h(\theta)|x-h(\theta)\rangle /2],
\ee
where $\theta$ is a physical parameter set including a chirp mass and a symmetric mass ratio, etc. 
In the above, $\langle x|h \rangle$ means the noise-weighted inner product given by 
\be \label{eq:innber_product}
\langle x|h \rangle = 2 \int_{f_{\rm min}}^{f_{\rm max}} \frac{\tilde{x}^* (f) \tilde{h}(f) +\tilde{x}(f) \tilde{h}^* (f) }{S_n (f)} df,
\ee
where $f_{\rm min}$ and $f_{\rm max}$ are the minimum and the maximum cutoff frequencies, respectively, and $S_n (f)$ is the noise power spectral density of the detector.

In the high SNR limit, the likelihood can be expressed as \cite{Cutler94, Cho13}
\bea
\nonumber L(\theta) &\propto& \exp [-\rho^2 \{1-\langle \hat{h}_0| \hat{h}(\theta)\rangle \}] \\
&=&  {\exp}[- \Gamma_{ij} \Delta \theta_i  \Delta \theta_j /2],
\eea
where $\hat{h}$ denotes the normalized waveform, $\hat{h}\equiv h/\sqrt{\langle h|h \rangle}$, 
$\rho$ is the SNR, $\rho\equiv \sqrt{\langle h_0|h_0 \rangle}$, $h_0$ is the GW signal,  and $\Gamma_{ij}$ is the FM defined by
\be \label{eq:fisher_matrix}
\Gamma_{ij} = \bigg \langle \frac{\partial h}{\partial \theta_i} \bigg| \frac{\partial h}{\partial \theta_j} \bigg \rangle \bigg|_{\theta_i=\theta_{\rm true}}.
\ee
The inverse of the FM corresponds to the covariance matrix of parameter errors. 
Thus, the measurement error $\sigma_i$ and the correlation coefficient $c_{ij}$ can be determined by using
\be \label{eq:error_and_correlation}
\sigma_i = \sqrt{(\Gamma^{-1})_{ii}}, \\
{c}_{ij} = \frac{(\Gamma^{-1})_{ij}}{\sqrt{(\Gamma^{-1})_{ii} (\Gamma^{-1})_{jj}}}.
\ee

%=================   section 3: result	====================
\section{Result}
We adopt a nonspinning equal-mass binary NS with $m_1 = m_2 = 1.4 M_\odot$ 
and choose $\Lambda_1 = \Lambda_2 = 300$ following the result in \cite{GW170817EOS}.
We consider the Advanced LIGO noise curve given in  \cite{Ajith09} and assume $f_{\rm min}$ = 10 Hz.
By using the TaylorF2 waveform, we calculate the 5$\times$5 FM 
whose components are $\theta_{i}$=\{${M_c}, \eta, \tilde{\Lambda}, {t_c}, {\phi_c} \}$ 
and present the measurement errors $\sigma_{M_c}, \sigma_{\eta}$, and $\sigma_{\tilde{\Lambda}}$.

\begin{figure}[h]
\begin{center}
\includegraphics[width=\columnwidth]{fig1.pdf}
\caption{\label{Fig-phase} The cumulative phase due to the tidal terms $\psi^{\rm Tidal}_{\rm 5PN}$ and $\psi^{\rm Tidal}_{\rm 5,6PN}$ (upper panel) and their fractional difference $(\psi^{\rm Tidal}_{\rm 5,6PN}-\psi^{\rm Tidal}_{\rm 5PN})/\psi^{\rm Tidal}_{\rm 5PN}$ (lower panel). We assume $f_{\rm min}=10$.}
\end{center}
\end{figure}

\begin{figure}[h]
\begin{center}
\includegraphics[width=\columnwidth]{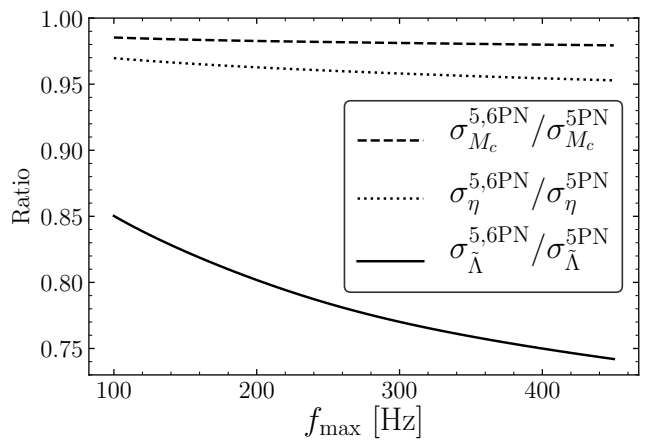}
\caption{\label{Fig-Error_ratio} Ratio of the measurement errors between the $\psi^{\rm Tidal}_{\rm 5PN}$ and the $\psi^{\rm Tidal}_{\rm 5,6PN}$ waveform cases (see text). We adopt a nonspinning equal-mass binary NS with $m_1 = m_2 = 1.4 M_\odot$ and $\Lambda_1 = \Lambda_2 = 300$. We consider the Advanced LIGO noise curve and assume $f_{\rm min}$ = 10 Hz. Note that the measurement error of $\tilde{\Lambda}$ can be reduced to about $75 \%$ at $450$ Hz by taking into account the tidal correction up to 6 PN
compared to the case in which only the 5 PN tidal correction is considered.}
\end{center}
\end{figure}

Figure 1 shows the cumulative phases from $10$ Hz to $f_{\rm max}$ 
due to the 5 PN tidal term ($\psi^{\rm Tidal}_{\rm 5PN}$) and
the sum of the 5 PN and the 6 PN tidal terms ($\psi^{\rm Tidal}_{\rm 5,6PN}$), respectively.
The fractional difference ($(\psi^{\rm Tidal}_{\rm 5,6PN}-\psi^{\rm Tidal}_{\rm 5PN})/\psi^{\rm Tidal}_{\rm 5PN}$)
is also given in Figure 1. 
We show the result only up to 450 Hz because our TaylorF2 model is fairly valid to 450 Hz \cite{Hinderer10}. 
Note that the fractional difference is about $20 \%$ at $450$ Hz. 

Next, we investigate how the 6 PN tidal term contributes to improving the accuracy in measuring the tidal parameter  $\tilde{\Lambda}$.
We take into account $\psi^{\rm Tidal}_{\rm 5PN}$ and $\psi^{\rm Tidal}_{\rm 5,6PN}$ in Eq. (\ref{eq:phase_of_TaylorF2})
and calculate the measurement errors for both cases.
Figure 2 shows the ratio of the measurement errors between the  $\psi^{\rm Tidal}_{\rm 5PN}$ and the $\psi^{\rm Tidal}_{\rm 5,6PN}$ waveform cases.
One can see that the measurement accuracy of the tidal deformability is significantly improved by including the 6 PN tidal correction. 
In this case, the measurement error of $\tilde{\Lambda}$ can be reduced to about $75 \%$ at $450$ Hz
compared to the case in which only the 5 PN tidal correction is considered.
On the other hand, the measurements of the mass parameters ($M_c$ and $\eta$) are almost not affected by the 6 PN tidal correction 
because that is mainly governed by the 3.5 PN  point-particle contribution ($\Psi_{\rm 3.5PN}^{\rm PP}$), 
which is about $10^{5}$ times larger than 
the tidal contribution ($\Psi^{\rm Tidal}$) in the cumulative phase evaluated from $10$ to $450$ Hz.

%=================   section 4: summary and discussion====================
\section{Summary and Discussion }\label{sec5}
We used  the TaylorF2 waveform model that incorporates the 3.5 PN order point-particle contribution with the tidal corrections.
By applying the FM method, we calculated the measurement errors of the mass and the tidal parameters for a nonspinning equal-mass binary NS. 
We found that the contribution of the 6 PN tidal correction to the cumulative phase from $10$ to $450$ Hz is about $20 \%$ of the contribution of the 5 PN tidal correction
and that the measurement error of the tidal deformability $\tilde{\Lambda}$ can be reduced to about $75 \%$ by including the 6 PN correction in the waveforms.

Tidal deformability is directly related with the NS EOS.
Therefore, in order to restrict various theoretical EOS models, 
one must reduce the measurement error of the tidal deformability.
In this context, we briefly demonstrated the necessity of using higher-order tidal corrections in the GW parameter estimation. 
 We assumed the maximum frequency cutoff to be 450 Hz in the overlap integration due to the validity of the waveform model.
 The changes in our result will be very small even if we consider much higher cutoff frequencies because 
 the sensitivity curve of Advanced LIGO used in this work increases very rapidly around $450$ Hz.
However, future third-generation GW detectors such as Einstein telescope \cite{Pun10,Hil11,Sat12} are much more sensitive than Advanced LIGO, especially in the high-frequency region;
therefore, the impact of the higher-order tidal corrections on the  accuracy in measuring the tidal deformability can be significantly 
increased by extending the cutoff frequency beyond 450 Hz.

%%%%%%%%%%%%%%%%%%%%%%%%%%%%%%%%%%%%%%%%%%%%%%%%%%%%%%%%%%%

%=======	Acknowledgements ===========================	
%

\section*{ACKNOWLEDGMENTS}
This work was supported by the National Research Foundation of Korea (NRF) grants funded by the 
Korea government (MSIP and MOE) (No. 2016R1A5A1013277, and No. 2018R1D1A1B07048599). 
HSC was supported by the National Research Foundation of Korea (NRF) grant funded by the Korea government (MSIP) (No. 2016R1C1B2010064).
%
%
%=======	Bibliography     ===========================	
%


\begin{thebibliography}{9}

\bibitem{Einstein1916} A. Einstein, Ann. Phys. {\bf 345}, 769 (1916).
\bibitem{Einstein1918} A. Einstein, Sitzungsber. K. Preuss. Akad. Wiss. {\bf 1}, 154 (1918).
\bibitem{Abbott16} B. P. Abbott \etal  (LIGO Scientific Collaboration and Virgo Collaboration), {\prl} {\bf 116}, 061102 (2016). %GW150914
\bibitem{Abbott16_2} B. P. Abbott \etal (LIGO Scientific Collaboration and Virgo Collaboration), {\prl} {\bf 116}, 241103 (2016). %GW151226
\bibitem{Abbott17} B. P. Abbott \etal (LIGO Scientific Collaboration and Virgo Collaboration), {\prl} {\bf 118}, 221101 (2017). %GW170104
\bibitem{Abbott17_2} B. P. Abbott \etal (LIGO Scientific Collaboration and Virgo Collaboration),  {\apj} {\bf 851}, L35 (2017). %GW170608
\bibitem{Abbott17_3} B. P. Abbott \etal (LIGO Scientific Collaboration and Virgo Collaboration), {\prl} {\bf 119}, 141101 (2017). %GW170814
\bibitem{GWCatalog_1} B. P. Abbott \etal (LIGO Scientific Collaboration and Virgo Collaboration), arXiv:1811.12907 (2018). %GWcatalog
\bibitem{GWCatalog_2} B. P. Abbott \etal (LIGO Scientific Collaboration and Virgo Collaboration), https://doi.org/10.7935/82H3-HH23  (2018). %GWcatalog


\bibitem{Abbott17_4} B. P. Abbott \etal (LIGO Scientific Collaboration and Virgo Collaboration), {\prl} {\bf 119}, 161101 (2017). %GW170817
\bibitem{Abbott17_5} B. P. Abbott \etal, Astrophys. J. Lett, {\bf 848}, L12 (2017). % ë…
\bibitem{Somiya12} K. Somiya \etal (KAGRA Collaboration), Classical Quantum Gravity {\bf 29}, 124007 (2012). % KAGRA …
\bibitem{Akutsu17} T. Akutsu \etal (KAGRA Collaboration), arXiv:1710.04823 (2017). 
\bibitem{Dominik15} M. Dominik \etal, {\apj} {\bf 806}, 263 (2015).
\bibitem{GW170817EOS} B. P. Abbott \etal (LIGO Scientific Collaboration and Virgo Collaboration), {\prl} {\bf 121}, 161101 (2018). %GW170817EOS


\bibitem{Raymond12} V. Raymond, Ph.D. thesis, Northwestern University, 2012.
\bibitem{Buonanno09} A. Buonanno, B. R. Iyer, E. Ochsner, Y. Pan and B. S. Sathyaprakash, {\prd} {\bf 80}, 084043 (2009).…\bibitem{Flanagan08} \'E. \'E. Flanagan and T. Hinderer, {\prd} {\bf 77}, 021502 (2008).

\bibitem{Lackey12} B. D. Lackey, Ph.D. thesis, The University of Wisconsin-Milwaukee, 2012. 

\bibitem{Hinderer09} T. Hinderer, {\apj} {\bf 697}, 964 (2009).

\bibitem{Hinderer10} T. Hinderer, B. D. Lackey, R. N. Lang and J. S. Read, {\prd} {\bf 81}, 123016 (2010).
\bibitem{Lackey15} B. D. Lackey and L. Wade, {\prd} {\bf 91}, 043002 (2015).
\bibitem{Favata14} M. Favata, {\prl} {\bf 112}, 101101 (2011).

\bibitem{Cho14} H.-S. Cho and C.-H. Lee, Classical Quantum Gravity {\bf 31}, 235009 (2014).
\bibitem{Cho15-1} H.-S. Cho,  {J. Korean Phys. Soc.} {\bf 66}, 1637 (2015).

\bibitem{Vallisneri08} M. Vallisneri, {\prd} {\bf 77}, 042001 (2008).







\bibitem{Cutler94} C. Cutler and \'E. E. Flanagan, {\prd} {\bf 49}, 2658 (1994).
\bibitem{Finn92} L. S. Finn, {\prd} {\bf 46}, 5236 (1992).

\bibitem{Cho13} H.-S. Cho, E. Ochsner, R. {O}'Shaughnessy, C. Kim and C.-H. Lee, {\prd} {\bf 87}, 024004 (2013).


\bibitem{Ajith09} P. Ajith and S. Bose, {\prd} {\bf 79}, 084032 (2009).



\bibitem{Pun10} M. Punturo \etal, Classical Quantum Gravity {\bf 27}, 194002 (2010).  
\bibitem{Hil11} S. Hild \etal,  Classical Quantum Gravity {\bf 28}, 094013 (2011).
\bibitem{Sat12} B. Sathyaprakash \etal, Classical Quantum Gravity {\bf 29},124013 (2012).



\end{thebibliography}
\end{document}